\documentstyle[11pt,twoside,newpasp,psfig,epsf]{article}

\newcommand{\ergs}{\mbox{{\rm\thinspace erg\thinspace s$^{-1}$}}}
\newcommand{\msun}{M_{\odot}}
\newcommand{\gtap}{\mathrel{\hbox{\rlap{\lower.55ex \hbox {$\sim$}}
                   \kern-.3em \raise.4ex \hbox{$>$}}}}
\newcommand{\ltap}{\mathrel{\hbox{\rlap{\lower.55ex \hbox {$\sim$}}
                   \kern-.3em \raise.4ex \hbox{$<$}}}}

\begin{document}

\title{X-ray sources in globular clusters}

\author{Frank Verbunt}
\affil{Astronomical Institute, Utrecht University, Postbox 80.000,
   3507 TA Utrecht, The Netherlands; email verbunt@phys.uu.nl}

\begin{abstract}
The twelve bright ($L_{\rm x}>10^{36}$ erg\,s$^{-1}$)
X-ray sources in the globular clusters 
have lower luminosities than the brightest sources in the bulge of
our galaxy. The dim ($L_{\rm x}<10^{35}$ erg\,s$^{-1}$) X-ray sources  
in globular clusters reach higher
luminosities than the cataclysmic variables in the disk of our galaxy.
The first difference is a statistical fluke, as comparison with M~31
indicates.
The second difference is explained because the brightest of the dim
sources are not cataclysmic variables, but soft X-ray transients
in quiescence.
This article describes the BeppoSAX, ROSAT and first Chandra
observations leading to these conclusions.
\end{abstract}


\section{Introduction}

\begin{figure}[]
\centerline{\psfig{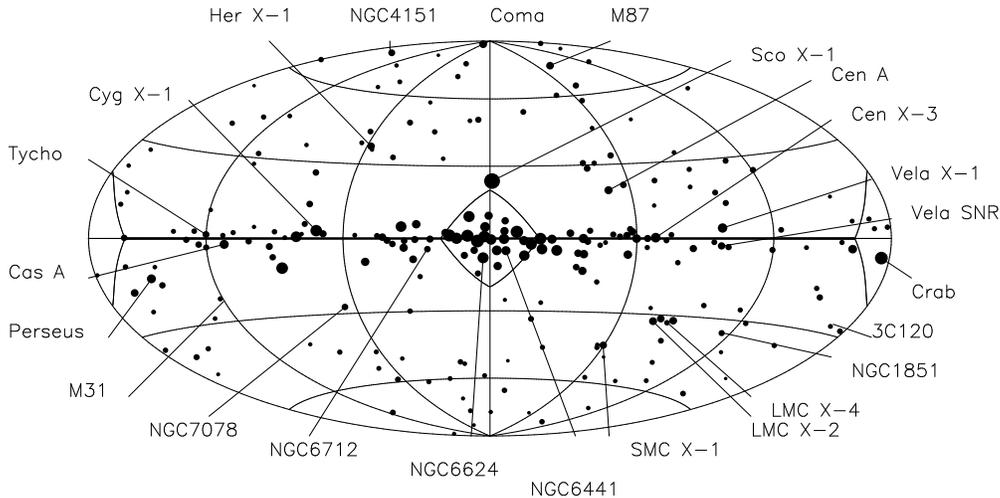}}
\caption{The brighter X-ray sources of the sky (from 
the 3rd Ariel\,V  catalogue, McHardy et al.\ 1981) are concentrated towards
the galactic plane, and in this towards the galactic center.
Thus a pointing at the galactic center with the
BeppoSAX Wide Field Cameras, with a field of view
indicated by the deformed lozenge in the middle,
observes half of the bright low-mass X-ray binaries
simultaneously, including many globular cluster
sources. Some globular clusters with bright X-ray sources are indicated
with their NGC numbers. \label{fvfiga}}
\end{figure}

In the 1970s the first maps of the X-ray sky showed that the
apparently brighter sources are concentrated towards the plane of the
galaxy, and in this plane towards the galactic center
(Figure\,\ref{fvfiga}). This indicates that these sources belong, in
majority, to our galaxy, and that they have distances large enough
that the galactic structure is evident, i.e.\ $\sim$8.5\,kpc.  
Most are binaries in which a neutron star or
black hole accretes matter from a companion star. The mass of the
donor star is the criterium to discriminate high-mass (donor
mass $\gtap 10\msun$) and low-mass ($\ltap 2\msun$) X-ray binaries
(as reviewed by e.g.\ Verbunt 1993). 
The luminosities of the low-mass X-ray binaries ranges up to
$L_x\simeq10^{38}\ergs$. Many are bright permanently,
but transient sources only during outbursts
which last a month, typically, and which repeat on time scales
ranging from six months to many decades (e.g.\ Chen et al.\ 1997).

Whereas globular clusters contain only of order 0.1\%\ of the stars 
in our galaxy, the current count is that they harbour 12 of the $\sim 100$
permanently or transiently bright low-mass X-ray sources.
The overabundance of bright X-ray sources in globular clusters
is caused by close encounters between neutron stars and main-sequence
or (sub)giant stars leading to the formation of low-mass X-ray
binaries (reviewed by Hut et al.\ 1992). 
A tidal capture occurs when a neutron star passes close 
enough to another star to raise a tidal bulge on it; the energy of
the bulge is taken from the relative orbit of the stars, which as a consequence
becomes bound. The efficiency of tidal capture is the subject of debate,
as rapid dissipation of the energy in the bulge may destroy the
star, leaving the neutron star alone again, or with a disk around it.
An exchange encounter is the process in which a neutron star takes
the place of one of the two members of a binary, by ejecting it.
The efficiency of exchange encounters depends on the number and
period distribution of binaries in the cluster cores. 

In addition to the bright sources, globular clusters also contain
dim sources, with X-ray luminosities $L_x\ltap10^{35}\ergs$, as discovered
with the Einstein satellite (Hertz \&\ Grindlay 1983). The nature of
these sources is less clear.

In this article I first explain how Jan van Paradijs awoke my
observational interest in cluster sources, and then review the
new results on the bright sources, mainly due to BeppoSAX; 
and the ROSAT and first Chandra results on the dim sources.

\begin{figure}[]
\centerline{\psfig{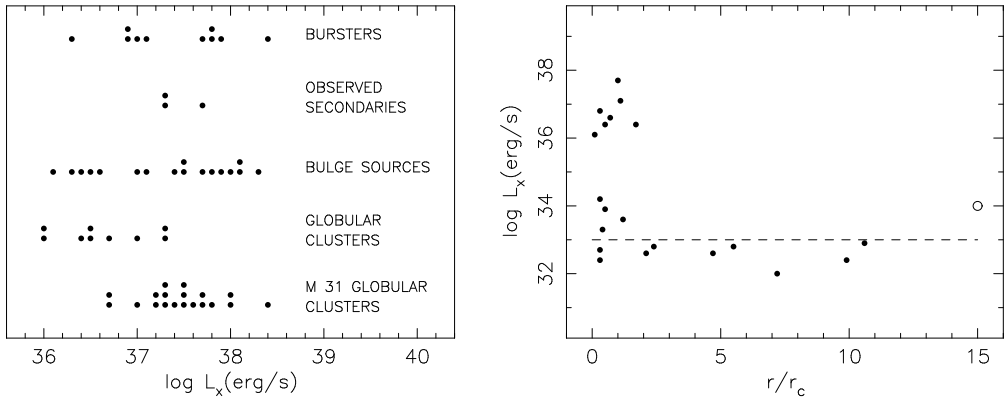}}
\caption{Reproduction of Figs.1 (left) and 4 (right) of Verbunt et 
al.\ (1984). The left figure showed that the globular cluster sources
in our galaxy do not reach the brighter luminosities measured for 
sources in the galactic disk (with distances from location near the galactic
center, from an optical companion, and from bursts), but that those in
globular clusters in M\,31 do reach these brightnesses.
The right figure shows that the dim ($<10^{35}\ergs$) 
globular cluster sources that are
brighter than the brightest cataclysmic variables in the galactic disk 
(i.e.\ above the dashed line) are within 2 core radii from the cluster
center, just as the bright cluster sources.
The brightest out-of-core source ($\circ$) is a marginal Einstein
source outside NGC\,5824, not detected in later, more sensitive
observations with ROSAT (Verbunt 2001).
Note that both figures would look different with current data.
\label{fvfigbc}}
\end{figure}

\section{Jan van Paradijs gets me involved}

In July 1983, as a postdoc in Cambridge (U.K.), I received a letter from Jan
van Paradijs in which he points out that the bright X-ray sources in
globular clusters do not reach the luminosities of their counterparts
near the galactic center (see Figure\,\ref{fvfigbc}); he asked me
whether it could be a consequence of lower mass donors in globular
clusters, due to a different formation mechanism in clusters (via
close encounters) than in the disk (via binary evolution).
While thinking about this, we ran into two puzzling facts (Verbunt et al.\ 
1984).

First, the X-ray sources in globular clusters in M\,31 {\em are} as
bright as the galactic bulge sources. This defeated any explanation we
could think of for the low luminosities of the cluster sources in our
galaxy, and I have since come to think that these low luminosities are
a statistical fluke.

Secondly, the dim sources had been suggested by their discoverers,
Hertz \&\ Grindlay, to be cataclysmic variables; however, they
are {\em brighter} in X-rays 
than the cataclysmic variables in the galactic disk.
Could they be neutron stars accreting at a low rate, i.e.\ soft
X-ray transients in their quiescent state? If so, they
should reside near the cores, and so we plotted the X-ray luminosities
of the globular cluster sources as a function of distance to the
cluster center, to find that the dim sources with $L_x\gtap10^{33}\ergs$
do indeed lie inside two core radii, just as the
bright sources (Fig.\,\ref{fvfigbc}). 
We thus proposed that the sources with $L_x\gtap10^{33}\ergs$ are
transients in quiescence, and those less luminous a mix of transients
and cataclysmic variables.
(Later it would appear that the Einstein sources outside the
core are not related to globular clusters -- see Sect.\,4.)

At the time of our proposal no transient had ever been detected in
X-rays in quiescence. To amend this, we obtained observations with
EXOSAT of two transients, Aql X-1 and Cen X-4. We detected Cen
X-4, and its luminosity was exactly in the range of brightest of the
dim globular cluster sources (Van Paradijs et al.\ 1987). 
This convinced us of the correctness of our proposal.

\section{New results on bright cluster sources}

The globular clusters are concentrated towards the galactic center,
and the $40^{\circ}\times40^{\circ}$ field of view of the BeppoSAX
Wide Field Cameras can observe a large fraction of them
simultaneously (Fig.\,\ref{fvfiga}), allowing the
discovery of rare events.

In a campaign of the galactic center, X-ray bursts -- thermonuclear
flashes on the surface of a neutron star -- have been discovered
in the bright sources of NGC\,6652 and
NGC\,6440 (In 't Zand et al.\ 1998, 1999).
These results bring the total of bright globular cluster sources
from which a burst has been detected to 11; leaving the source
in Terzan\,6 as the  only one remaining
in which no thermonuclear burst has been unambiguously detected.
This means that 11 of the 12 currently known permanent or transient 
bright sources in globular clusters, and possibly all 12 of them, 
harbour a neutron star. 
(It may be remarked that one of the other elusive burst sources,
the one in M\,15, was caught by Van Paradijs et al.\ 1990; I thank
Phil Charles for reminding me of this.)

A possible explanation for the absence of X-ray binaries with a
black hole in globular clusters has been provided by Portegies Zwart \&
McMillan (2000; see also Kulkarni et al.\ 1993). 
Once the more massive stars in a globular cluster have evolved,
the black holes -- if they have masses similar to the black holes
in the galactic disk, i.e.\ of about 7\,$\msun$ (Bailyn et al.\ 
1998) -- are the most massive objects in the cluster by a large
margin, the next most massive stars being the 1.4$\,\msun$ neutron
stars. As a result the black holes sink to the deepest part of the potential
well, where they form binaries by exchanging into primordial binaries. 
Interactions between binary
and single black holes lead to reaction velocities larger than
the escape velocity of the cluster, and effectively clear the
globular cluster of black holes.

Another result of BeppoSAX was the discovery of faint, short-lived 
X-ray transients, e.g.\ in NGC\,6440 
(in 't Zand et al.\ 1999). This raises the question how many of
such transients are missed by other satellites, that do not
have the spatial resolution to overcome source confusion and
discover such outbursts in the vicinity of the galactic center.
The outburst in NGC\,6440 of 1998 reached a maximum of only $6\,10^{36}\ergs$,
and had an exponential decay time of about 5\,d;
very different from the 1972/1973 outburst, which stayed at a luminosity
of $3\,10^{37}\ergs$ during more than a month (Markert et al.\ 1975,
Forman et al.\ 1975).
This raises the question whether NGC\,6440 harbours more than
one transient; or alternatively how one source can show such extremely
different outbursts.

Finally, with BeppoSAX eclipses were discovered in the source in Terzan\,6,
and on the basis of these the orbital period was determined to be 12.36\,h
(in 't Zand et al.\ 2000). 
Of the five orbital periods now known for globular cluster sources,
only one has a period in the range 
compatible with a main-sequence donor, viz.\ the source
in NGC\,6441 with a period of about 5.7\,h (Sansom et al.\ 1993).
Two have periods suggesting a subgiant donor, viz.\ 12.36 and 17.1\,h;
and two periods suggesting a white-dwarf donor, at 11.4 and 20.6\,m
(Ilovaisky et al.\ 1993, Stella et al.\ 1987, Homer et al.\ 1996).
Van Paradijs \&\ McClintock (1994) found an observational relation between the
X-ray luminosity, the orbital period, and the visual brightness of
low-mass X-ray binaries. They also provided a heuristic argument to explain
it: a large orbital period allows a large disk,
which when irradiated by a high X-ray luminosity will be a bright
source of optial radiation. Conversely, if the optical counterpart of a
bright X-ray source is faint in the optical, the orbital period must
be small, and this reasoning is used by Deutsch  et al.\ (2000) and Homer et 
al.\ (2001)
to argue that the donors of the sources in NGC\,1851 and NGC\,6652 also
are white dwarfs.
The prevalence of ultra-short period systems, and thus presumably
of white-dwarf donors among the globular cluster X-ray sources underscores
the difference in formation mechanism of globular cluster X-ray sources
and of low-mass X-ray binaries in the galactic disk.
In particular, it indicates the importance in clusters of direct collisions
between single neutron stars and sub-giants in the formation of
ultra-compact binaries (Verbunt 1987).

\section{The dim sources: ROSAT}

\begin{figure}[]
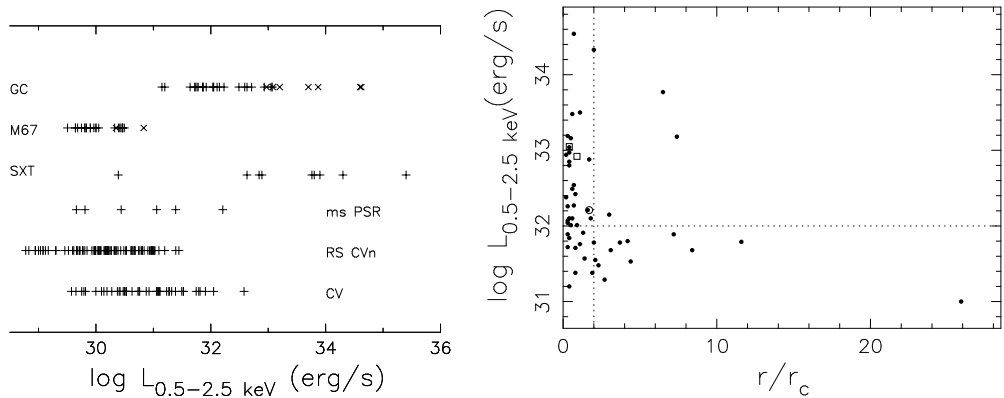

\centerline{
\parbox[b]{6.0cm}{\psfig{figure=verbuntf3a.ps,width=6.0cm,clip=t}}
\parbox[b]{7.2cm}{\psfig{figure=verbuntf3b.ps,width=7.2cm,clip=t}}
}
\caption{Left: The X-ray luminosities as determined with ROSAT
of the dim sources in globular clusters
compared to those of various types of sources in the galactic
disk, from top to bottom: sources in the old open clusters M~67 and NGC~188,
soft X-ray transients, recycled millisecond pulsars, RS CVn binaries,
and cataclysmic variables. For each category, the lower
limit is set by the detection limit.
The luminosities of the brightest dim sources
are matched by the soft X-ray transients only
(from Verbunt 1996). Right:
X-ray luminosities of globular cluster sources as a function
of their distance to the cluster center, in units of the core
radius, based on ROSAT data (Verbunt (2001). The proposed 
quiescent soft X-ray transient in $\omega$\,Cen is indicated
$\circ$, those in 47\,Tuc $\Box$. The horizontal dotted line indicates
the level of the brightest cataclysmic variables in the galactic disk,
which due to the narrower bandpass is lower than the (Einstein) level
of Fig.\,\ref{fvfigbc}.
\label{fvfigde}}
\end{figure}

The debate on the nature of the dim sources widened as ROSAT
observations of sources in the galactic disk showed that 
recycled neutron stars and RS CVn binaries may have luminosities
comparable to the luminosities of the faintest dim sources.
ROSAT observations of cataclysmic variables and soft X-ray
transients in quiescence confirmed that only the transients
reach the luminosities of the brightest dim sources
(Figure\,\ref{fvfigde}). ROSAT also showed that soft
X-ray transients in quiescence show a soft spectrum
($kT_{\rm bb}\ltap0.4$\,keV for a black body fit;
Verbunt et al.\ 1994), to which ASCA added
a hard power-law tail (Asai et al.\ 1998). 
In an important paper, Brown et al.\ (1998) 
showed that the quiescent X-ray spectrum of soft X-ray
transients is quiescence is dominated by thermal emission
from the neutron star, and that fits of neutron-star 
atmosphere models leads to realistic radii for the neutron stars.
The few dim cluster sources for which ROSAT could provide
spectral information appear to be equally soft (Verbunt 2001).

It further became clear that the dim sources found with
Einstein outside the cluster cores (and shown in Fig.\,\ref{fvfigbc})
are fore- or background sources (Margon \&\ Bolte 1987,
Cool et al.\ 1995), or nonexistent (Verbunt 2001)!
However, ROSAT detected new out-of-the core sources which according
to a simple statistical argument in majority {\em are} related to the
globular clusters (Verbunt 2001); this includes a dwarf
nova in the outskirts of NGC\,5904 (Hakala et al.\ 1997).

The ROSAT HRI managed to
resolve the dim sources in the cores of 47\,Tuc, NGC\,6397 and NGC\,6752
in 5, 4 or 5, and 4 sources, respectively; in $\omega$\,Cen
three sources were detected in the core (Verbunt \&\
Hasinger 1998, Verbunt \&\ Johnston 2000).
Second, the more accurate ROSAT positions spurred HST 
searches for optical counterparts.
Several blue variables -- possibly cataclysmic variables --
in the core of 47\,Tuc were proposed as counterparts to the
X-ray sources (Paresce et al.\ 1992, Paresce \&\ de Marchi
1994, Shara et al.\ 1996); but chance coincidence could
not be excluded (Verbunt \&\ Hasinger 1998).
Similarly, the proposed identifications of stars with
H\,$\alpha$ emission in NGC\,6397 (Grindlay et al.\ 1995, Edmonds et al.\ 
1999), $\omega$\,Cen (Carson et al.\ 2000),
and of variables with periods of 5.1 and 3.7\,h in NGC\,6752 (Bailyn
et al.\ 1996), with dim X-ray sources in these clusters are plausible,
but due to the uncertain positions not certain.
In my view it has been too easily assumed that these
sources are cataclysmic variables: variability and
H\,$\alpha$ emission are also properties of X-ray
transients in quiescence. In any case, none of the proposed
identifications concern the brighter dim sources proposed 
to be quiescent soft X-ray transients by Verbunt et al.\ (1984).
The one secure identification of a dim core source before Chandra is
that based on the periodicity of the recycled pulsar in M\,28 (not by
Danner et al.\ 1994, who unwittingly analyzed background photons, but
by Saito et al.\ 1997; see discussion in Verbunt 2001).
The luminosity function of the core sources was analyzed by
Johnston \&\ Verbunt (1996), who found that the total
luminosity of most globular clusters is dominated by
a few relatively bright dim sources, rather than large numbers of 
very dim sources.

Finally, a comparison showed that most globular cluster have a lower
X-ray luminosity to mass ratio than the old open cluster M\,67 
(observed by Belloni et al.\ 1998), which indicates that
sources of the (ill-understood) types responsible for most X-rays 
in M\,67 are not present in globular clusters (Verbunt 2001).

\section{The first Chandra results on dim sources}

\begin{figure}[]
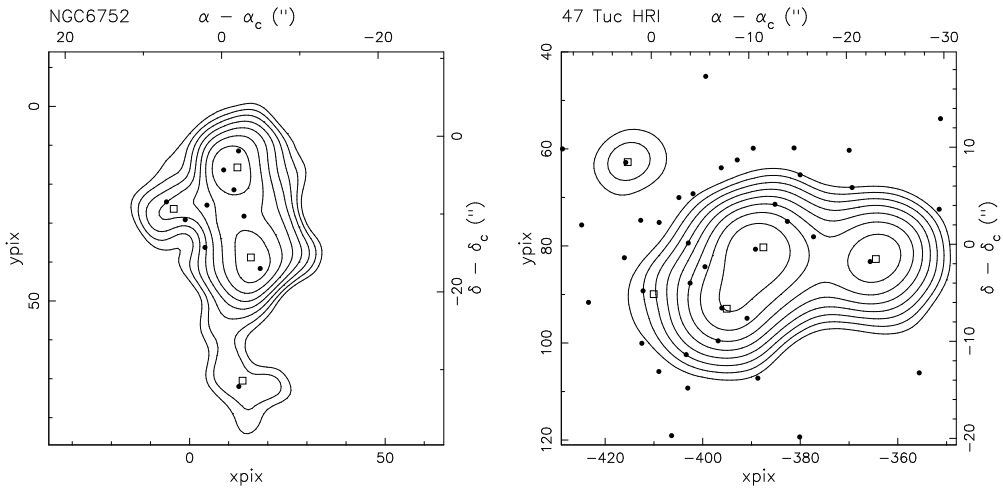

\centerline{
\parbox[b]{6.6cm}{\psfig{figure=verbuntf4a.ps,width=6.6cm,clip=t}}
\parbox[b]{6.6cm}{\psfig{figure=verbuntf4b.ps,width=6.6cm,clip=t}}
}
\caption{The X-ray positions of sources in cluster cores detected with
Chandra ($\bullet$) superposed on ROSAT HRI contours and positions ($\Box$).
Left NGC\,6752 (Pooley et al.\ 2001; Verbunt \&\ Johnston
2000); right 47\,Tuc (Grindlay et al.\ 2001; Verbunt \&\ Hasinger 1998).
\label{fvfigfg}}
\end{figure}

The first images of globular clusters obtained with Chandra
confirm its superior angular resolution as compared to the
already impressive ROSAT HRI.
Chandra resolves the central source of NGC\,6752
into nine sources (Pooley et al.\ 2001; see 
Fig.\,\ref{fvfigfg}).
Five of these sources have X-ray fluxes and colours 
suggestive of cataclysmic variables.
Work on the optical identification is in progress, and has
found evidence for H\,$\alpha$ emission for the candidate
counterparts of these sources (Homer et al.\ in preparation).
However, the flux distribution of the
brightest X-ray source is best fitted with a power law.
The two optical  variables discovered by Bailyn et al.\ (see
previous section) are both detected as X-ray sources.

The globular cluster $\omega$\,Cen has a larger core, and 
Chandra doesn't add to the three sources already found with
ROSAT in this core. However, the Chandra observation does provide
spectral information, and this has been used by Rutledge et al.\
(2001) to show that the brightest source in $\omega$\,Cen,
outside the cluster core but within the half-mass radius,
has a spectrum similar to those of soft X-ray transients in
the galactic disk. The spectral fit gives a radius of about 10\,km
at the distance of $\omega$\,Cen, and confirms the 
membership of this source to the cluster.

The most spectacular Chandra observation of a globular cluster
is that of 47\,Tuc, in which a large number of sources is detected
(Grindlay et al.\ 2001).
On the basis of their soft X-ray colours, 
the brightest two sources (X\,5 and X\,7 from the ROSAT HRI 
observations; Verbunt \&\ Hasinger 1998) are argued to be
soft X-ray transients in quiescence -- a nice confirmation
of the suggestion discussed in Sect.\,2.
Thirteen sources are probably cataclysmic variables, on the
basis of their hard X-ray colours.
Fifteen recycled radio pulsars have been found with the 
source detection algorithm;
the 1$''$ accuracy of the Chandra positions allows the
detections of two of them on the basis of {\em three} photons
and {\em one} photon,  respectively!
Six sources can be optically identified with main-sequence
binaries.

\begin{figure}[]
\centerline{\psfig{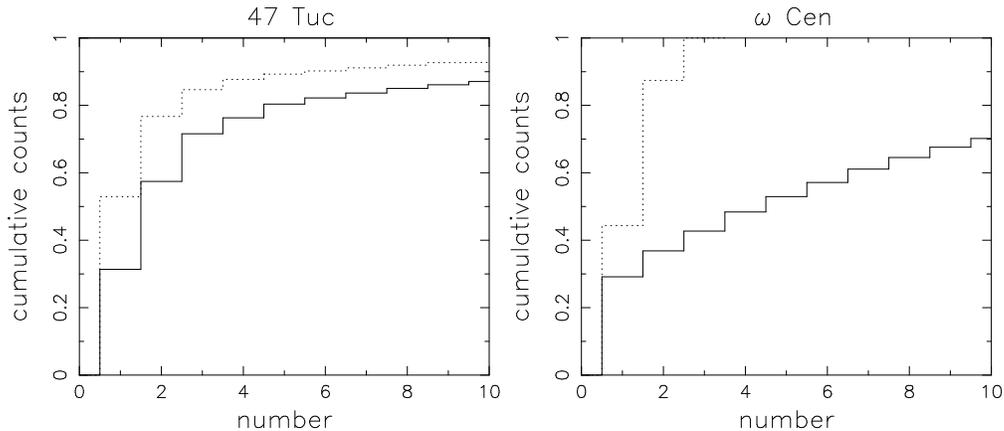}}
\caption{Cumulative fraction of total X-ray luminosity partaken by
sources detected with Chandra in 47\,Tuc (left) and in $\omega$\,Cen.
The full line is for all the sources near the cluster, including
background sources; the dotted
line for the core only.
\label{fvfigh}}
\end{figure}

In Fig.\,\ref{fvfigfg} I plot the positions of the Chandra
sources in 47~Tuc on top of the ROSAT HRI image. Comparison of the positions
of 8 HRI sources and their Chandra counterparts, shows that
the overall shift required to bring the ROSAT positions into
agreement with the Chandra positions is $-0\farcs7,-0\farcs5$
in right ascension and declination respectively; after application
of this shift, all 8 sources coincide within the errors with
the Chandra sources. Thus the positional accuracy claimed by
Verbunt \&\ Hasinger is confirmed.
In Fig.\,\ref{fvfigh} I plot the cumulative X-ray flux 
of the Chandra sources, starting with the brightest one.
This plot confirms the ROSAT result that the total luminosity of
47\,Tuc is dominated by the few brightest sources.
For example, the fifteen detected recycled pulsars together emit
less than 4\%\ of the flux of the brightest source alone.

The measured period derivatives of the recycled pulsars are
dominated by gravitational acceleration in the cluster
(Freire et al.\ 2001). The low X-ray luminosities of these
pulsars provide an indirect estimate to their intrinsic
period derivatives, and thus to their age, if we use the
empirical relation found by Verbunt et al.\ (1996) that
$L_{\rm x}\simeq10^{-3}L_{\rm sd}$, where $L_{\rm sd}=4\pi^2I\dot P/P^3$
is the spindown luminosity.
If we take the average period of the recycled pulsars in 47\,Tuc 
as $\sim$4\,ms, and the average X-ray luminosity
$L_{\rm x}\sim3\,10^{30}\ergs$, the characteristic age
$\tau_c\equiv P/(2\dot P)$ is comparable to the age of the cluster.
This suggests that most currently detected recycled pulsars were
formed soon after the formation of 47\,Tuc.

\section{Conclusions and prospects}

The 12 permanently or transient bright X-ray sources in the globular clusters
of our galaxy are neutron stars. The dearth of high-luminosity
($\gtap10^{37}\ergs$, say) sources among them appears to
be a fluke of small-number statistics.

Among the dim sources, all types proposed as possible dim sources,
i.e.\ soft X-ray transients in quiescence, cataclysmic variables,
recycled pulsars, and chromospherically active close binaries
are now detected in Chandra observations.
For more secure identification of H\,$\alpha$-emission systems
with cataclysmic variables, it would be useful if a compilation
of H\,$\alpha$ line fluxes of soft X-ray transients {\em in
quiescence} were available.
The total core luminosity of a globular cluster is dominated by a
few bright sources, rather than large number of dim sources.
Globular cluster cores and {\em a fortiori} globular
clusters as a whole emit less X-rays per unit mass than
the old open cluster M\,67.

Chandra has opened a golden age for the study of X-ray sources
in globular clusters. Together with HST, this will for the first
time provide us with good statistics of the numbers of binaries
in the cores of globular clusters, an important ingredient in the
study of cluster evolution.

\end{document}